# Water Waves from General, Time-Dependent Surface Pressure Distribution in the Presence of a Shear Current

Yan Li[*], Simen Å. Ellingsen
Department of Energy and Process Engineering, Norwegian University of Science and Technology,
Trondheim, Norway

**We obtain a general solution for the water waves resulting from a general, time-dependent surface pressure distribution, in the presence of a shear current of uniform vorticity beneath the surface, in three dimensions. Linearized governing equations and boundary conditions including the effects of gravity, a distributed external pressure disturbance, and constant finite depth, are solved analytically, and particular attention is paid to classic initial value problems: an initial pressure impulse and a steady pressure distribution which appears suddenly. In the present paper, good agreements with previous results are demonstrated. We subsequently show both analytically and numerically how transient waves from a suddenly appearing steady pressure distribution vanish for large times, and steady ship waves remain. The transient contribution to wave resistance was derived. The results show that a shear current has significant impact on the transient wave motions, resulting in asymmetry between upstream and downstream waves. The case of the suddenly appearing steady pressure is an intermediate case between ring waves and ship waves, starting out as the former and evolving gradually into the latter. The ship's direction of motion relative to the current is found to be crucial in determining how quickly effects of transient waves die out. Transient effects take much longer to die out for a ship going against the shear flow than for one going downstream, when the two ships have the same velocity relative to the water surface. For ship motion against the shear current, thus, wave effects of transients, e.g., due to maneuvering could accumulate and be far more significant than on a uniform current.**

INTRODUCTION

Surface waves in the presence of sub-surface shear currents are both physically fascinating and of practical importance. Most work on this topic, both analytical and numerical, has been performed in two dimensions, where much progress has been made (e.g., Peregrine, 1976; Teles da Silva & Peregrine, 1988). Of particular interest has been the simplest shear flow model, where vorticity is assumed to be spatially constant. In two dimensions this flow still permits the use of potential theory for linear waves, and the constant vorticity means that complications from critical layers are avoided (Booker & Bretherton, 1967), although only in absence of time-varying submerged sources (Ellingsen & Tyvand, 2015a). The literature on this particular flow is extensive; see, e.g. Ellingsen & Brevik (2014) and references therein. In three dimensions, however, progress on the interaction of waves and shear flow is more recent. Ellingsen (2014b) considered ship waves in the presence of shear currents, and also the classical Cauchy-Poisson problem (2014a) with initially prescribed surface shape and velocity. Li & Ellingsen (2015a) extended work of Ellingsen (2014b) to finite waters, in which interactions of waves, shear current and seabed are fully discussed. Progress was also recently made on the classical problem of a submerged oscillating point source in 3D by Ellingsen & Tyvand (2015b) with a shear current present.

Classically, two different types of initial value problems have been considered for surface waves. The Cauchy-Poisson variant is one, in which the surface shape and vertical velocity are prescribed at the initial time. An alternative is to consider waves generated by applying a localized pressure impulse when the water is initially at rest (Stoker, 1957). When the pressure impulse is short (a Dirac delta function in time), the initial conditions may be reformulated in terms of an initial velocity distribution, and the problem is equivalent to a Cauchy-Poisson problem with prescribed velocity. A formalism in terms of a general transient external pressure is useful, however, in the study of ship waves, where a "ship" is typically modelled as a travelling pressure source depressing the surface (see Darmon et al. 2014; Ellingsen, 2014b; Raphaël & Gennes, 1996). The general formalism includes, in principle, unsteady or accelerated ship motion, and is thus a step towards a more general consideration of non-stationary ship waves in the presence of a shear current.

Physically, a steady motion must be the limiting behaviour of a transient process starting at $t=0$ (say) and tending to a steady state as $t \rightarrow \infty$. It is illustrated by Stoker (1957) that the radiation condition, which is imposed to avoid unphysical waves "from infinity", is not necessary if a steady state is reached as an asymptote of an initial value problem. In general, researchers tend to focus on the steady state, which is actually a somewhat artificial problem since every steady motion must have commenced in the finite past, see Stoker (1957); Darmon et al. (2014); Li & Ellingsen (2015b). A few, however, have described the passage from unsteady motion to the steady ship waves formulated by means of a surface pressure (Stoker 1953, Wurtele 1955), and the sibling problem of waves resulting from the sudden appearance of submerged sources has been studied (see Wehausen & Laitone, 1960-§22). As a further example of the formalism given in the present paper, we formulate waves from a suddenly appearing moving pressure source. This thus provides a perspective which shows the passage from the unsteady to the steady state in the presence of a shear current.

The presence of a shear current can drastically affect wave motion, resulting in a significantly anisotropic dispersion relation. This results in a widening, narrowing or skewing of ship waves (Ellingsen, 2014b), and ring waves from a localized source become elongated and, when shear is strong, cease being ring-shaped at all (Ellingsen, 2014a). In the present paper, a general expression is first derived for initial conditions as well as a general, transient external pressure at the free surface. Secondly, the related patterns with initial pressure impulse are obtained. For this particular case, a numerical study of wave patterns and how they are affected by finite depth, vorticity and surface current velocity is presented. Finally, waves and wave resistance from a suddenly appearing steady pressure distribution are investigated by applying the general solution obtained. The waves in the latter case can be divided into a transient and a stationary contribution, the latter being the steady ship wave pattern found previously (Ellingsen 2014b, Li & Ellingsen 2015a), while the transient contribution is shown analytically to vanish as a function of time in the manner of a ring wave. For this case, a numerical study of wave patterns which shows how transient waves evolve into steady waves is presented. The transient contribution to the wave resistance is calculated, adding an oscillatory contribution to the wave resistance. While the transient vanishes for large times, the time it takes for transient waves to have vanished can vary greatly for different directions of motion. For ship motion against the shear current, transient effects can take a long time to vanish, indicating the possibility that transient waves, for example due to maneuvering, can more easily accumulate and have significant effects on the ship's motion than in a uniform current.

## FORMULATION AND GENERAL SOLUTIONS

### Description of the Problem
We consider the wave-current system depicted in Figure 1. It is assumed that the water is incompressible, and the water viscosity is negligible. Fluid motion here is rotational due to the shear current, and three dimensional, thus potential theory cannot be used (Ellingsen & Brevik, 2014). We consider water with constant depth $h$.

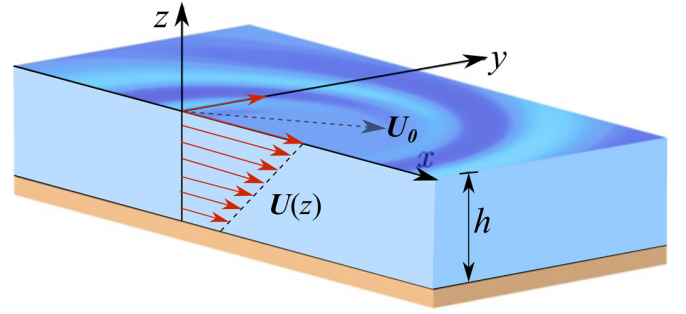

Figure 1. Three dimensional wave-current and coordinate systems

**Governing equations and linearized boundary conditions**
The governing equation for the fluid field is the continuity equation and Euler equation

$$\nabla \cdot \mathbf{V} = 0 ; \tag{1}$$

$$\frac{D\mathbf{V}}{Dt} = -\frac{1}{\rho}\nabla P + \mathbf{g}, \tag{2}$$

in which operator $\nabla = (\partial/\partial x, \partial/\partial y, \partial/\partial z)$; $\mathbf{V} = \mathbf{V}(x, y, z, t) = (U_x + u, U_y + v, w)$, is the fluid velocity, with the basic shear flow $\mathbf{U}(z) = (U_x, U_y) = (Sz + U_{0x}, U_{0y})$ having constant vorticity $S$, and the small velocity perturbations are $u, v, w$; The total pressure is $P(x, y, z, t)$ and the small pressure perturbation is $p$ so that $P = -\rho g z + p$. The density of the fluid is $\rho$; and $\mathbf{g}$ is the acceleration of gravity pointing in the $-z$ direction.

Boundary conditions include the free surface and sea bottom conditions. The linearized kinematic and dynamic boundary conditions at the free surface are expressed as

$$\left. \begin{aligned} w|_{z=0} &= \frac{\partial \zeta}{\partial t} + \mathbf{U}_0 \cdot \nabla \zeta \\ (P - \rho g \zeta)|_{z=0} &= p_{ext} \end{aligned} \right\}, \tag{3}$$

where $\zeta = \zeta(\mathbf{r}, t)$ is the surface elevation compared to the undisturbed surface, and $p_{ext}(\mathbf{r}, t)$ is the external pressure disturbance acting as a wave source, which we assume known. Everywhere, $\mathbf{r} = x\vec{i} + y\vec{j}$.
We furthermore assume that the external pressure is zero for $t<0$ and has arbitrary time dependence thereafter, but in such a way that its temporal Laplace transform exists. At the sea bottom, the boundary condition is
$w|_{z=-h} = 0$. (4)

When viscosity is neglected, the flow is fully determined by Eqs. 1 ~ 4 and initial conditions.

### Fourier Transformation
We introduce Fourier transformations in the $xy$ plane, defined as

$$\left. \begin{aligned} [u,v,w](\mathbf{r},z,t) &= \iint \frac{d^2\mathbf{k}}{(2\pi)^2} [\tilde{u}, \tilde{v}, \tilde{w}](\mathbf{k},z,t) e^{i\mathbf{k}\cdot\mathbf{r}}, \\ p(\mathbf{r},z,t) &= \iint \frac{d^2\mathbf{k}}{(2\pi)^2} \tilde{p}(\mathbf{k},z,t) e^{i\mathbf{k}\cdot\mathbf{r}}, \\ \zeta(\mathbf{r},t) &= \iint \frac{d^2\mathbf{k}}{(2\pi)^2} \tilde{\zeta}(\mathbf{k},t) e^{i\mathbf{k}\cdot\mathbf{r}}, \\ p_{ext}(\mathbf{r},t) &= \iint \frac{d^2\mathbf{k}}{(2\pi)^2} \tilde{p}_{ext}(\mathbf{k},t) e^{i\mathbf{k}\cdot\mathbf{r}}, \end{aligned} \right\} \tag{5}$$

in which wave vector $\mathbf{k} = (k_x, k_y) = (k\cos\theta, k\sin\theta)$.

Linearize Eq. 1 and Eq. 2, and apply Fourier transformation to Eq. 1~4, then the continuity and Euler equations turn to:

$$\left.\begin{array}{l}ik_x\tilde{u}+ik_y\tilde{v}+\tilde{w}'=0, \quad \dot{\tilde{u}}+i\mathbf{k}\cdot\mathbf{U}\tilde{u}+S\tilde{w}=-ik_x\tilde{p}/\rho,\\ \dot{\tilde{v}}+i\mathbf{k}\cdot\mathbf{U}\tilde{v}=-ik_y\tilde{p}/\rho, \quad \dot{\tilde{w}}+i\mathbf{k}\cdot\mathbf{U}\tilde{w}=-\tilde{p}'/\rho\end{array}\right\} \quad (6)$$

here, a dot represents derivation with respect to time, a prime with respect to $z$.

### General Solutions in Finite Water Depth

Following the procedure of Ellingsen (2014a) by eliminating $\tilde{u}$, $\tilde{v}$, and $\tilde{p}$, we obtain a Rayleigh equation for $\tilde{w}$ alone,

$$(\partial/\partial t + i\mathbf{k}\cdot\mathbf{U})(\partial^2/\partial z^2 - k^2)\tilde{w} = 0. \quad (7)$$

Based on Eq. 7, the general solution of $\tilde{w}$ has the expression:

$$\tilde{w} = kA(\mathbf{k},t)\sinh k(z+h) + kC(\mathbf{k},t)\cosh k(z+h) + \frac{k^3 D(\mathbf{k})e^{-ik_x Ut}}{k^2 + (k_x St)^2} \quad (8)$$

in which $A(\mathbf{k}, t)$, $C(\mathbf{k}, t)$ and $D(\mathbf{k})$ are spatially constant. Applying the seabed boundary condition into Eq. 8 yields $C(\mathbf{k},t) = D(\mathbf{k}) = 0$.

Substituting $\tilde{w}$ back into the equations, we obtain

$$\left.\begin{array}{l}\tilde{w} = kA(\mathbf{k},t)\sinh k(z+h)\\ \tilde{p}/\rho = -(\dot{A} + i\mathbf{k}\cdot\mathbf{U}A)\cosh k(z+h) + (iSk_x A/k)\sinh k(z+h)\end{array}\right\}. \quad (9)$$

Substituting Eq. 9 into the boundary conditions at the free surface yields

$$\left.\begin{array}{l}kA(\mathbf{k},t)\sinh kh = \dot{\tilde{\zeta}} + i\mathbf{k}\cdot\mathbf{U}\tilde{\zeta}\\ -(\dot{A} + i\mathbf{k}\cdot\mathbf{U}A)\cosh kh + (iSk_x A/k)\sinh kh - g\tilde{\zeta} = \tilde{p}_{ext}/\rho\end{array}\right\}, \quad (10)$$

in which $U_0$ is the current velocity at the still free surface. For later reference we define the shorthand $A_0 = A(\mathbf{k},0)$ and $\tilde{\zeta}_0 = \tilde{\zeta}(\mathbf{k},0)$.

We apply a Laplace transform to Eq. 10 and eliminate $A$ to get

$$\overline{\tilde{\zeta}}(\mathbf{k},s) = \overline{f_{I_{ext}}}(\mathbf{k},s)/(s^2 + 2i\omega_1 s + \omega_2^2), \quad (11)$$

where the Laplace transform is defined as

$$\overline{f}(s) = \int_0^\infty dt\, f(t)e^{-st}, \quad (12)$$

and we have defined the quantities

$$\left.\begin{array}{l}\omega_1 = \mathbf{k}\cdot\mathbf{U}_0 - Sk_x \tanh kh/(2k),\\ \omega_2^2 = (k_x S\mathbf{k}\cdot\mathbf{U}_0/k + gk)\tanh kh - (\mathbf{k}\cdot\mathbf{U})^2,\\ \overline{f_{I_{ext}}}(\mathbf{k},s) = -(k\overline{\tilde{p}_{ext}}(\mathbf{k},s)\tanh kh)/\rho + k\sinh kh A_0\\ \quad + [s + i\mathbf{k}\cdot\mathbf{U} - (iSk_x/k)\tanh kh]\tilde{\zeta}_0\end{array}\right\}. \quad (13)$$

Hence the free surface elevation could be expressed as

$$\zeta(x,y,t) = \iint \frac{d^2k}{(2\pi)^2}\tilde{\zeta}(\mathbf{k},t)e^{i\mathbf{k}\cdot\mathbf{r}}, \quad (14)$$

where $\tilde{\zeta}(\mathbf{k},t) = \int_\Gamma \frac{ds}{2\pi i}\overline{f_{I_{ext}}}(\mathbf{k},s)e^{st}/(s^2 + 2i\omega_1 s + \omega_2^2)$, and the contour $\Gamma$ runs from $-i\infty$ to $+i\infty$ to the right of all singularities

After we get the expression of the surface elevation, we can obtain the vertical velocity distributions by Eq. 9. Eqs. 11-14 denote general solutions of surface elevation resulting from a time dependent applied pressure disturbance, including Cauchy-Poisson problems (refer to Ellingsen 2014(a) for detailed information), and, as a limiting case, ship waves.

## INITIAL IMPULSIVE PRESSURE

We shall consider the classical initial condition of an impulsive pressure applied to the surface when the water is initially at rest, which just remains for an infinitesimally short time at $t=0$, but imparts a finite momentum to the surface. Such a pulse is described by a Dirac delta function. In this paper, we focus on this case as well as a steady pressure source within the fully general framework developed above.

Considering time $t > 0$. The time-dependent pressure disturbance has returned to zero, which means the first term of $\overline{f_{I_{ext}}}(\mathbf{k},s)$ in Eq. 13 is zero. Now performing the inverse Laplace transform gives

$$\zeta(x,y,t) = \iint \frac{d^2k}{(2\pi)^2}\left\{[kA_0\sinh kh - iSk_x\tilde{\zeta}_0\tanh kh/(2k)]\times\right.$$
$$\left.\frac{\sin\sqrt{\omega_1^2+\omega_2^2}\,t}{\sqrt{\omega_1^2+\omega_2^2}} + \tilde{\zeta}_0\cos\sqrt{\omega_1^2+\omega_2^2}\,t\right\}e^{i(\mathbf{k}\cdot\mathbf{r}-\omega_1 t)} \quad (15)$$

Equation 15 describes a pressure impulse problem as well as a Cauchy-Poisson problem. In the latter case, $\zeta(x,y,t)$ and $\partial\zeta/\partial t$ are given at $t=0$, and $p_{ext}(x,y,t)=0$. With a slight change of formalism, Eq. 13 and Eq. 15 could be expressed as

$$\zeta(x,y,t) = \iint \frac{d^2k}{(2\pi)^2}\left\{(i\omega_1\tilde{\zeta}_0 + \dot{\tilde{\zeta}}_0)\frac{\sin\sqrt{\omega_1^2+\omega_2^2}\,t}{\sqrt{\omega_1^2+\omega_2^2}}\right.$$
$$\left.+ \tilde{\zeta}_0\cos\sqrt{\omega_1^2+\omega_2^2}\,t\right\}e^{i(\mathbf{k}\cdot\mathbf{r}-\omega_1 t)} \quad (16)$$

which is identical to Eq. 25 in Ellingsen (2014a) when using the conventions defined therein.

For the second initial case, according to Eq. 15, $A_0$ and $\tilde{\zeta}_0$ have yet to be determined. For easy comparison with literature (e.g. Darmon et al (2014)) a Gaussian distribution is used to define the initial pressure impulse, which we express as

$$p_I(r) = I\delta(t)e^{-(\pi r/a)^2}, \quad (17)$$

wherein $a$ is the width of the pulse, and

$$\left.\begin{array}{l}\tilde{p}_I = I\delta(t)a^2 e^{-(ka/2\pi)^2}/\pi\\ \int_{-\infty}^{+\infty}\delta(t)dt = \int_{0^-}^{0^+}\delta(t)dt = 1\end{array}\right\}. \quad (18)$$

Next, we obtain the value of $A_0$ and $\tilde{\zeta}_0$ by integrating Eq. 10 over $t$ in the interval from $0^-$ to $0^+$, whereby

$$\left.\begin{array}{l}\int_{0^-}^{0^+}dt\, kA\sinh kh = \int_{0^-}^{0^+}dt\, i\mathbf{k}\cdot\mathbf{U}_0\widetilde{\zeta_{Ip}} + \int_{0^-}^{0^+}dt\,\frac{\partial\widetilde{\zeta_{Ip}}}{\partial t}\\ \int_{0^-}^{0^+}dt\,\{-(\dot{A}+i\mathbf{k}\cdot\mathbf{U}_0 A)\cosh kh + (iSk_x A/k)\sinh kh - g\widetilde{\zeta_{Ip}}\} = \int_{0^-}^{0^+}dt\,\tilde{p}_I/\rho\end{array}\right\}$$

(19)

in which $\widetilde{\zeta_{Ip}}$ denotes the surface elevation resulting from an initial impulsive pressure.

By the assumption that all perturbation quantities are zero when $t<0$, $A$ and $\zeta$ must take finite values at $t = 0^+$. Hence, from Eq. 19, we obtain

$$\tilde{\zeta}_0 = 0, \quad A_0 = -\frac{a^2 I e^{-(ka/2\pi)^2}}{\pi \rho \cosh kh}. \quad (20)$$

With these values, the surface elevation can be expressed as

$$\zeta(x,y,t) = \iint \frac{d^2k}{(2\pi)^2} \left\{ -\frac{ka^2 I e^{-(ka/2\pi)^2}}{\rho\pi} \tanh kh \frac{\sin\sqrt{\omega_1^2+\omega_2^2}\,t}{\sqrt{\omega_1^2+\omega_2^2}} e^{i(\mathbf{k}\cdot\mathbf{r}-\omega_1 t)} \right\}. \quad (21)$$

Table 1. Physical V.S. dimensionless quantities

| Physical quantities | Non-dimensional quantities |
|---|---|
| $\zeta$ | $\zeta/a$ |
| $h$ | $H=h/a$ |
| $r$ ($\mathbf{r}$) | $R=r/a$ ($\mathbf{R}=\mathbf{r}/a$) |
| $(x,y)$ | $(X,Y) = (x/a, y/a)$ |
| $t$ | $T = t/\sqrt{a/g}$ |
| $\omega_{1,2}$ | $\Omega_{1,2} = \omega_{1,2}\sqrt{a/g}$ |
| $k$ ($\mathbf{k}$) | $K=ak$ ($\mathbf{K}=a\mathbf{k}$) |
| $S$ | $Fr_S = S\sqrt{a/g}$ |
| $U_0$ | $Fr = \frac{U_0}{\sqrt{ga}}$ ; $Fr_{SU_0} = \frac{SU_0}{g}$ ; $Fr_h = \frac{U_0}{\sqrt{gh}}$ |
| $I$ | $PI = I/(\rho a \sqrt{ag})$ |

In order to better evaluate effects of velocity, vorticity and water depth, we introduce the non-dimensional quantities using similar rescaling rules as Ellingsen (2014a), defined in Table. 1. $Fr$ and $Fr_S$ are "surface current Froude number" and "shear Froude number", respectively, based on the current velocity at the surface $U_0$ and velocity $bS$, $Fr_h$ is the Froude number based on water depth and used to estimate the seabed effects. In terms of non-dimensional quantities, the surface elevation now takes the form

$$\frac{\zeta}{a} = \int_{-\pi}^{\pi} d\gamma \int_0^{\infty} \frac{-K^2 e^{-\left(\frac{K}{2\pi}\right)^2} PI \tanh KH \sin T \sqrt{(Fr_s \cos\theta \tanh KH)^2 + K \tanh KH}}{4\pi^3 \sqrt{(Fr_s \cos\theta \tanh KH)^2 + K \tanh KH}}$$
$$e^{i(\mathbf{K}\cdot\mathbf{R} - (KFr\cos\gamma - Fr_s \cos\theta \tanh KH)T)} dK \quad (22)$$

## STEADY PRESSURE DISTRIBUTION AT FREE SURFACE

Owing to the ability to superpose solutions to a linear problem, the result from an initial pressure impulse can be used to obtain the solution for a suddenly appearing steady pressure distribution, provided that the pressures have the same spatial distribution (Cummins, 1962). Let us therefore use the Gaussian distribution also to simulate the suddenly appearing steady pressure,

$$p_{ext}(r) = p_0 e^{-(\pi r/a)^2} \quad (23)$$

with Fourier transform

$$\widetilde{p_{ext}} = p_0 a^2 e^{-(ka/2\pi)^2} / \pi \quad (24)$$

Using a Gaussian shaped disturbance to model a "ship" has the benefit of introducing only a single parameter, $a$, into our already parameter rich model. The resulting ship waves will share many traits with those produced by a more realistic ship hull, although differences can be significant in certain Froude number regimes (e.g., He et al, 2015). The model can be generalized to an oblong, ellipsoidal Gaussian in a very straightforward manner (Benzaquen et al 2014), and the general expressions introduced above can readily accommodate a more realistic modelling of the applied pressure to mimic a realistic traveling wave source.

The resulting surface elevation at time $t > 0$ can be obtained by summing the contributions from a continuous string of pressure impulses since $t=0$,

$$\zeta_{sp}(x,y,t) = \int_0^t dt_0 \zeta_{Ip}(x,y,t-t_0), \quad (25)$$

Substituting Eq. (21) in Eq. (25) yields

$$\zeta_{sp}(x,y,t) = \zeta^s(x,y) + \zeta^t(x,y,t) \quad (26)$$

in which

$$\zeta^s(x,y) = -\iint \frac{d^2k}{(2\pi)^2} \frac{\widetilde{p_{ext}} k \tanh kh}{\rho \omega_2^2} e^{i\mathbf{k}\cdot\mathbf{r}}$$

$$\zeta^t(x,y,t) = \iint \frac{d^2k}{(2\pi)^2} \frac{\widetilde{p_{ext}} k \tanh kh}{2\rho \sqrt{\omega_1^2+\omega_2^2}} \left(\frac{e^{-i\omega_+ t}}{\omega_+} - \frac{e^{-i\omega_- t}}{\omega_-}\right) e^{i\mathbf{k}\cdot\mathbf{r}} \quad (27)$$

with

$$\omega_{\pm} = \omega_1 \pm \sqrt{\omega_1^2+\omega_2^2}$$
$$\omega_2^2 = -\omega_+ \omega_- \quad (28)$$

We see that the surface elevation expression $\zeta_{sp}$ splits naturally into two main parts: the time-independent $\zeta^s$ and time-dependent $\zeta^t$. Comparing with the ship wave expression with finite depth and shear current in Li & Ellingsen (2015a) [or Ellingsen (2015b) for infinite depth] shows that $\zeta^s$ exactly equals the steady ship wave result but for a difference of 180º in the definition of $\beta$ due to different choice of coordinate system. Full details on how the steady ship wave is calculated numerically and discussions of the particulars of the resulting patterns are provided by Li & Ellingsen (2015a) and need not be iterated here. Our focus here is to show how the time-dependent term $\zeta^t$ is transient and will, ultimately, vanish for large times as intuitively it must.

Alternatively, response from a steady pressure disturbance can also be obtained by Eq. 14 with zero initial impulsive pressure and surface elevation. Not surprisingly, the result is the same with Eqs. 25-28.

### *Integrand poles*

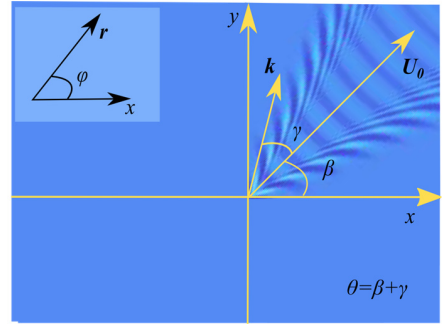

Figure 2. Definition of the angles

In order to evaluate the transient behavior of the wave pattern at large times, we must consider the various poles in the integrands of Eq. 27. Following the analysis of Stoker (1957) we here discuss the possible roots of the denominator of (27), bearing in mind that since $\omega_2^2 = -\omega_+\omega_-$, this will also include the poles of equation (26). The dispersion relation for the static ship waves is

$\omega_2^2 = (g + SU_0 \cos\theta\cos\gamma)k \tanh kh - k^2 U_0^2 \cos^2\gamma = 0$

(definitions of $\gamma$ and $\theta$ are shown in Fig.2), which is equivalent with

$$\left.\begin{array}{l}\tanh(KH) = KH/B \\ B = \dfrac{1 + Fr_{SU_0}\cos\theta\cos\gamma}{Fr_h^2\cos^2\gamma}\end{array}\right\} \qquad (29)$$

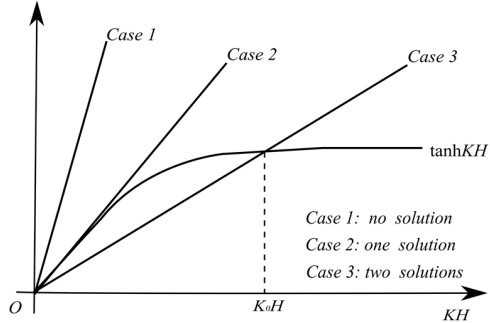

Figure 3. Possible Solutions of Eq. 29. Graphs show $KH/B$ for three different values of $B$.

Possible solutions of Eq. 29 are shown in Fig.3, including three distinct situations based on value of $B$. Clearly, only real roots will contribute to real physical wave oscillations, and thus imaginary roots (which contribute only to near-field solutions) will not be considered. The three different cases are:

*Case (1)*: $B<1$, there is only one real root at $KH=0$, and it is a double root. Since the numerators of the fractions in the integrands of Eq. 27 also have a double zero at $k=0$, there is no corresponding pole in the integrands of Eq. 27.

*Case (2)*: $B=1$, a Taylor expansion in the vicinity of $k=0$ yields

$$\omega_2^2 = U_0^2 \cos(\gamma)^2(B-1)k^2 - \frac{h^3 k^4 g}{3}BFr_h^2\cos^2\gamma + O(k^6) \qquad (30)$$

Obviously, there is only one real root at $K=0$, and it is a quadruple root.

*Case (3)*: $B>1$, there are two real roots at $K=0$ (which does not give a pole in Eq. 27) and at a finite value $K=K_0>0$.

The following analysis will refer to these three different situations, in order to show how transient waves vanish with time while waves in the far field survive. We further assume that $U_0\neq 0$ for the analyses in this chapter, although the limit $U_0 \to 0$ is considered numerically.

*Radiation condition*
The splitting of Eq. 25 into two different terms in Eq. 26 has come at a price, because while equation 25 is well defined and can be evaluated directly, each of the terms in Eq. 26 has poles on the integration axis which make the integrals in Eq. 27 undetermined. Dealing with this specific problem, a radiation condition is necessary to ensure only waves radiating energy *away* from the source are included (e.g. Raphaël & Gennes (1996), Stoker (1957) and Mei (1989)). Various mathematical ways to formulate radiation conditions are possible, and the one which is most suitable for the present case is arguably the heuristic inclusion of a very small dissipation, ensuring that wave amplitudes decrease in time.

This is mathematically equivalent to the condition often used for ship wave problems where the ship is assumed to have been "slowly switched on" (Raphaël & Gennes (1996), Lighthill 1978, Li & Ellingsen 2015a). Multiplying all perturbation quantities by a dissipation factor $e^{\varepsilon t}$ ($\varepsilon>0$) from the outset, we find a solution which is also proportional to $e^{\varepsilon t}$, and finally take the limit of the solution as $\varepsilon \to 0$:

$$[p_{ext}, u, v, w] \to [p_{ext}, u, v, w]e^{\varepsilon t}, \qquad (31)$$

in which $\varepsilon$ is an infinitesimal positive number that will ultimately approach to zero. In the limit $\varepsilon \to 0$ the replacement rule 31 is only significant to the extent that it moves the position of the poles in the integrands of Eq. 27 away from the real axis, and hence the solutions in Eq. 26 - 27 are substituted by

$$\zeta_{sp}(\boldsymbol{r},t) = \lim_{\varepsilon \to 0^+}\left\{\zeta_\varepsilon^s(\boldsymbol{r}) + \zeta_\varepsilon^t(\boldsymbol{r},t)\right\}, \qquad (32)$$

in which

$$\left.\begin{array}{l}\zeta_\varepsilon^s(\boldsymbol{r}) = -\iint \dfrac{d^2k}{(2\pi)^2}\dfrac{\widetilde{p_{ext}}k\tanh kh}{\rho(\omega_2^2 + i\varepsilon(2\boldsymbol{k}\cdot\boldsymbol{U}_0 - S\cos\theta\tanh kh))}e^{i\boldsymbol{k}\cdot\boldsymbol{r}} \\ \zeta_\varepsilon^t(\boldsymbol{r},t) = \iint \dfrac{d^2k}{(2\pi)^2}\dfrac{\widetilde{p_{ext}}k\tanh kh}{2\rho\sqrt{\omega_1^2+\omega_2^2}}\left(\dfrac{e^{-i\omega_+ t}}{\omega_+ - i\varepsilon} - \dfrac{e^{-i\omega_- t}}{\omega_- - i\varepsilon}\right)e^{i\boldsymbol{k}\cdot\boldsymbol{r}}\end{array}\right\}. \qquad (33)$$

*Transient waves at large times*
We then discuss the asymptotic behavior of the time-dependent part for large $t$, rewriting $\zeta_\varepsilon^t(x,y,t)$ in the form

$$\zeta_\varepsilon^t(\boldsymbol{r},t) = \int_{-\pi}^{\pi}\frac{d\theta}{(2\pi)^2}I(\theta), \qquad (34)$$

in which

$$\left.\begin{array}{l}I(\theta) = I_+(\theta) + I_-(\theta) \\ I_+(\theta) = \int_0^\infty dk\, f(k)\dfrac{e^{-i\omega_+ t}}{\omega_+ - i\varepsilon} \\ I_-(\theta) = \int_0^\infty dk\, f(k)\dfrac{e^{-i\omega_- t}}{\omega_- - i\varepsilon} \\ f(k) = \dfrac{\widetilde{p_{ext}}k^2\tanh kh}{2\rho\sqrt{\omega_1^2+\omega_2^2}}e^{i\boldsymbol{k}\cdot\boldsymbol{r}}\end{array}\right\}. \qquad (35)$$

For *case (1)*, we first take the limit of $\varepsilon \to 0^+$ as there is no singularity in the integrand of Eq. 34 after taking the limit. Following Bender & Orszag (1999) the behaviour at large $t$ can be found by using a stationary phase method, which can be expressed

$$\lim_{t \to +\infty}\int_a^b dk\, f(k)e^{it\psi(k)} \sim i\sqrt{\frac{2\pi}{t|\psi''(k_0)|}}f(k_0)e^{it\psi(k_0)+i\text{sgn}[\psi''(k_0)]\pi/4}, \qquad (36a)$$

$$\lim_{t \to +\infty}\int_a^b dk\, f(k)e^{it\psi(k)} \sim O(t^{-1}), \qquad (36b)$$

in which $f(k)$ is a regular function in the interval $a<k<b$. Eq. 36a is relevant when a stationary point $k_0$ exists, so that $a<k_0<b$, $\psi'(k_0) = 0$, $f(k_0) \neq 0$ and $\psi''(k_0) < 0$, and if the stationary point is one of the end points, the asymptotic behavior is just half of the result. If there is no such point $k_0$ satisfying $\psi'(k_0) = 0$ in the interval, the asymptotic behavior is Eq. 36b.

In *case (1)*, the functions $\omega_+(k)$ and $\omega_-(k)=0$ have non-vanishing first derivatives for all $k$, and consequently $\zeta_{sp}$ will behave like $1/t$ as $t \to +\infty$ since there are no stationary point at which the phase is stationary.

*Case (2)*, is the special limit of *case (3)* when $k_0$ tends to zero. We do not dwell on the particular mathematical properties of this case since it is not associated with any new physical behaviour, but rather consider it a limit of *case (3)* which we consider below.

The most interesting case is *case (3)*. The integrand for Eq. 32 has no singularity at the origin $k=0$, but a simple pole at $k_0 + i\varepsilon/\omega'_\pm(k_0)$ ($k_0$ is a real number and larger than zero). The main function of $\varepsilon$ is to slightly move the singularity out of the integral path so that a proper mathematical method can be applied for asymptotic behaviors at large times and in far fields. The Cauchy residue theorem is used in the following analyses, which is expressed (Needham, 1997)

$$\oint_C dx f(x) = 2\pi i \sum_{j=1}^{n} Res\{f, a_j\} \quad (37)$$

in which $a_j$ is the singular point of $f(x)$ inside the closed contour $C$. Equivalently, an integral along path in the complex plane can be shifted freely so long as the end points remain the same and no singularities are crossed. If shifting the integral path means that a pole is crossed, the new integral differs from the old by a term which is the residual term in Eq (27).

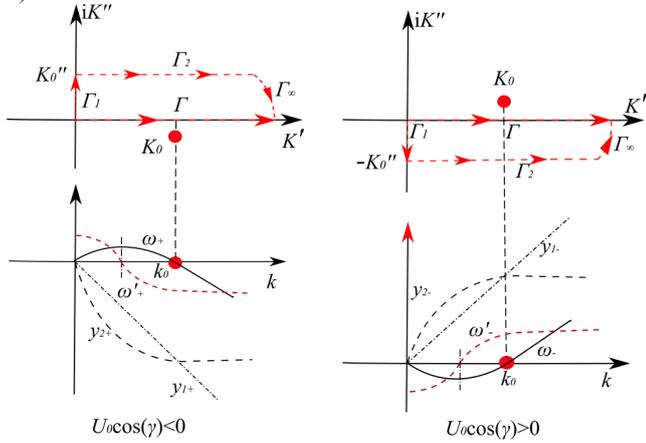

Figure 4. The integral paths and solutions in $k$ plane for two different situations: $U_0 \cos\gamma < 0$ and $U_0 \cos\gamma > 0$. In the figure, $\omega_\pm(k) = y_{1\pm} - y_{2\pm}(k)$ with $y_{1\pm}(k) = kU_0\cos\gamma$ and $y_{2\pm}(k) = \mp\sqrt{gk\tanh kh + \frac{S^2\cos^2\theta \tanh^2 kh}{4}} - \frac{S\cos\theta\tanh kh}{2}$.

Before analyzing the asymptotic behaviors at large times, we have to check roots of $\omega_\pm = 0$ first, shown in Fig. 4, because the possible roots might be singularities of integrand for Eq. 35. As illustrated in Fig. 4, there is a non-zero root for $\omega_+=0$ ($\omega_-=0$) if and only if $U_0\cos\gamma < 0$ ($U_0\cos\gamma > 0$). Additionally, there are various roots on the imaginary axis which contribute to the near-field only, and when shifting integral contours we must be careful not to cross these imaginary axis poles, the nearest of which is always a nonzero distance from the real axis.

We discuss only the case $U_0\cos\gamma < 0$ in detail since the analysis for $U_0\cos\gamma > 0$ is essentially identical. Note that $U_0$ is always larger than (or equal to) zero as it's the absolute value of current velocity at still water surface. The present analysis is quite similar to Stoker (1956) and Lamb (1932), in which the integral with singularities in integrand over the integral path between 0 and $+\infty$ is fully discussed. Moving the contour slightly above the real $k$ axis as shown in the top right panel of Fig. 4 we obtain

$$I_+(\theta) = \int_{\Gamma_1+\Gamma_2+\Gamma_\infty} dk\, f(k) \frac{e^{-i\omega_+ t}}{\omega_+ - i\varepsilon} \quad (38)$$

in which integral paths $\Gamma_1$, $\Gamma_2$, and $\Gamma_\infty$, are shown in the left-hand side of Fig. 4. We chose to move the contour above the real axis in order to ensure that the contribution from $\Gamma_\infty$ be exponentially small for large $t$. Since the pole near $k_0$ is above the axis, no pole was crossed in shifting the integral path. Note that we can always choose path $\Gamma_1$ short enough so that no imaginary root of $\omega_+=0$ lies on $\Gamma_1$. For large times, the integral over path $\Gamma_1$ will then fall off like $1/t$ as there is no stationary point; over path $\Gamma_2$ the single stationary point means the integral falls off like $1/\sqrt{t}$ as there is one point which satisfies $\omega'_+(k)=0$ while $\omega''_+(k)<0$. Over path $\Gamma_\infty$ the integral over $\Gamma_\infty$ is exponentially small when $t$ is large. We have thus demonstrated that $I_+(\theta)$ vanishes for large $t$ when $U_0\cos\gamma < 0$ as $1/\sqrt{t}$. No restrictive assumptions are necessary for this argument other than assuming the pressure distribution $\widetilde{p_{ext}}(k)$ is analytical on or inside the closed contour and $f(k)$ having limited value along $\Gamma_\infty$.

When still assuming $U_0\cos\gamma < 0$, integral $I_-(\theta)$ has no stationary points along its original (real) integration path, hence both $I_+(\theta)$ and $I_-(\theta)$ vanish for large $t$ when $U_0\cos\gamma < 0$.

The case $U_0\cos\gamma > 0$, can be analyzed in identical manner except that we must now shift the path into the lower complex $k$ plane in order for $\Gamma_\infty$ to vanish, while the pole near $k_0$ now appears in the upper half of the complex plane. The conclusion is the same, that the time-dependent part of the surface elevation will vanish for large times.

*Steady waves & the special case with Fr=0*
The previous sections have illustrated that the transient waves will ultimately vanish for large times, and the steady wave remains. The time independent term in Eq. 26 describes the ship waves resulting from a moving pressure source of Gaussian shape, and is the same as Eq. 2.15 in Li & Ellingsen (2015a) when using the conventions defined therein. For an in-depth analysis regarding the steady ship wave pattern in the presence of a shear current, refer to Ellingsen (2014b) and Li & Ellingsen (2015a, b).

When *Fr*=0, which is to say that the pressure distribution is at rest relative to the free surface, we can demonstrate, again, that the transient waves would, ultimately, vanish at large times.

## LIMITING CASES AND NUMERICAL EXAMPLES

*Dispersion Relation*
From Eqs. 10-13, we can obtain the dispersion relation which could be expressed as

$$\left[\omega - \mathbf{k}\cdot\mathbf{U}_0 + \frac{Sk_x\tanh kh}{2k}\right]^2 = gk\tanh kh + (Sk_x\tanh kh)^2/(4k^2), \quad (39)$$

in which $\omega$ is the absolute frequency. This relation reduces to the well known case (e.g., Mei (1989)) when the current is uniform (i.e. $S$=0). When there is no current at all, Eq. 39 is obviously the classical dispersion relation $\omega^2 = gk\tanh kh$.

The intrinsic frequency $\sigma$ could be defined in the expression as

$$\sigma = \omega - \mathbf{k}\cdot\mathbf{U}_0 = \pm\sqrt{gk\tanh kh + \left(\frac{Sk_x\tanh kh}{2k}\right)^2} - \frac{Sk_x\tanh kh}{2k} \quad (40)$$

In this perspective, we can define the group velocity relative to the

surface velocity

$$\boldsymbol{C}_g = \nabla_k \sigma = \frac{\partial \sigma}{\partial k_x}\mathbf{i} + \frac{\partial \sigma}{\partial k_y}\mathbf{j} = \frac{\partial \sigma}{\partial k}\mathbf{e}_k + \frac{1}{k}\frac{\partial \sigma}{\partial \theta}\mathbf{e}_\theta \quad (41)$$

where $\mathbf{e}_k$ and $\mathbf{e}_\theta$ are unit vectors in polar coordinate Fourier space.

*Effect of an initial pressure impulse*

From Eq. 20 we see that the initial pressure impulse at time $t=0^+$ directly results in an initial surface velocity distribution given by $A_0$, subsequently developing into an outward propagating surface wave.

Fig.5 depicts the vertical velocity distributions for given initial pressure impulse. It may be seen in Fig. 5 that, at $t = 0^+$, the main vertical velocity disturbance caused is focused in the region $r \leq a$. Outside this region, the vertical velocity is relatively small. It is obvious that the initial vertical velocity will predominately have the opposite sign to the initial pressure impulse.

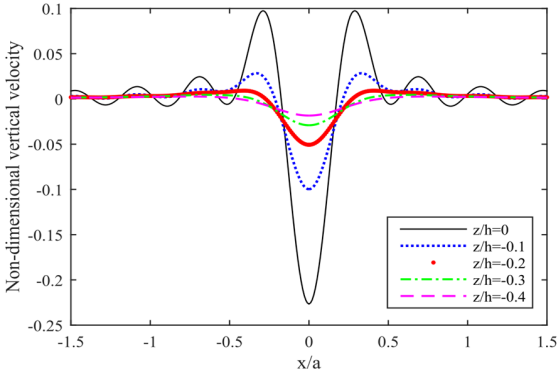

Figure 5. Non-dimensional vertical velocity distributions $W = w/\sqrt{ga}$ at $t=0$ as $H=20$, $PI=$ 2e-4.

*The Effect of Water Depth*

*Solutions in infinite water depth*
By taking the deep water limit of solutions in finite water depth, we can obtain the specific solutions for infinite water depth, which can be expressed as

$$\left.\begin{array}{l}\tilde{w} = kAe^{kz} \\ \tilde{p}/\rho = -[\dot{A} + ik_x A(U + S/k)]e^{kz} + Const \\ \tilde{\zeta}(\boldsymbol{k},t) = \int_\Gamma \frac{ds}{2\pi i}\frac{-k\bar{\tilde{p}}_{ext}(\boldsymbol{k},s)/\rho + kA(\boldsymbol{k},0) + [s + ik_x(U_0 - S/k)]\tilde{\zeta}(\boldsymbol{k},0)}{(s^2 + 2i\omega_1 s + \omega_2^2)}e^{st}\end{array}\right\}$$

(42)

Here, $\omega_1 = \boldsymbol{k}\cdot\boldsymbol{U}_0 - Sk_x/(2k)$, and $\omega_2^2 = gk + k_x S\boldsymbol{k}\cdot\boldsymbol{U}_0/k - (\boldsymbol{k}\cdot\boldsymbol{U}_0)^2$.

For the corresponding Cauchy-Poisson initial value problem, we obtain the same expression of surface elevation as given in Eq. 16, with $\omega_1$ and $\omega_2$ defined as in Eq. 42.

For the second initial value problem, we can get the solution of surface elevation expressed as

$$\frac{\zeta(X,Y,T)}{a} = -\text{PI}\iint \frac{d^2K}{(2\pi)^2}K\frac{\sin\sqrt{\Omega_1^2 + \Omega_2^2}T}{\sqrt{\Omega_1^2 + \Omega_2^2}}e^{i(\boldsymbol{K}\cdot\boldsymbol{R}-\Omega_1 T)-(K/2\pi)^2}, \quad (43)$$

In which $\Omega_1$ and $\Omega_2$ are defined as before while with $\tanh KH$ approaching to 1. The exponential ensures that the main contribution comes from $K \lesssim 2\pi$.

Eq. 43 corresponds to Stoker (1957, p.159) when no current is included and assuming the same initial pressure distribution.

The dispersion relation in infinite water depth could be expressed as

$$\sigma = \omega - \boldsymbol{k}\cdot\boldsymbol{U}_0 = \pm\sqrt{gk + \left(\frac{S}{2}\cos\theta\right)^2} - \frac{S}{2}\cos\theta \quad . \quad (44)$$

According the eq. 44, we can get the relative group velocity along the direction of phase propagation expressed as

$$C_g = \frac{1}{2}\sqrt{\frac{g}{k}}\frac{1}{\sqrt{1 + (S\cos\theta)^2/(4k)}} < \frac{1}{2}\sqrt{\frac{g}{k}} \quad . \quad (45)$$

And the relative phase velocity $c$ could be expressed as

$$c = \sigma/k = \sqrt{\frac{g}{k}}\sqrt{1 + \frac{1}{4gk}S^2\cos^2\theta} - \frac{S}{2k}\cos\theta \lessgtr \sqrt{\frac{g}{k}} \quad (46)$$

As discussed by Ellingsen (2014a) we may disregard the $\pm$ in Eq. 46 when considering the wave velocities and retain the positive value, since the two signs correspond to (positive) velocities in opposite directions, and there is a unique positive phase and group velocity in each direction $\theta$. Eqs. 45 and 46 reveal that the relative phase velocity is greater than the group velocity in every wave propagating direction, as one expects for gravity waves. Note moreover how the presence of the shear flow can increase or decrease the phase velocity, but only decrease the group velocity.

*Limit of shallow water*
We consider the shallow water situation, when $h \ll 1$. Then, $\tanh kh \sim kh$. From Eq. 39~Eq. 41 we obtain

$$C_g \approx c \approx \sqrt{gh + \left(\frac{Sh}{2}\cos\theta\right)^2} - \frac{Sh}{2}\cos\theta \quad . \quad (47)$$

Eq. 47 shows that the relative phase and group velocities coincide when the water depth is small compared to a wavelength, and there is no dispersion despite the sub-surface shear current. However, velocity varies with the direction of wave propagation due to the existence of shear current. The lack of dispersion means that wave patterns retain their ring shape as they propagate.

Fig. 6 shows wave patterns at deep ($H$=20) and shallow ($H$=0.05) waters at different times with moderate shear current. In the figure, a moderate shear $Fr_s$=1 is considered and the surface current velocity ($Fr_{U_0}$=0) is ignored. It is obvious from Fig.6 that the shear current has different influence upon waves in deep and shallow water. Effects of the shear current are more prominent in deep waters. This accords to Eq. 45~47. The effects of shear in shallow waters are diminished, and the lack of dispersion is apparent from the stability of the ring shapes as they propagate. These observations concur with those of Ellingsen (2014a).

The previous section has shown that the vorticity has different degree of influence in deep and shallow waters. Consider now strong, moderate and zero shear in deep waters, without the surface current velocity in order to simplify the problem. The situation is shown in Fig. 7. In Fig. 7, we can observe that a clear asymmetry is created by the shear current upon the waves from the symmetrical initial pressure disturbance. When the shear is strong enough, the waves fronts are no longer ring shaped, as can be seen in Fig. 7-c.

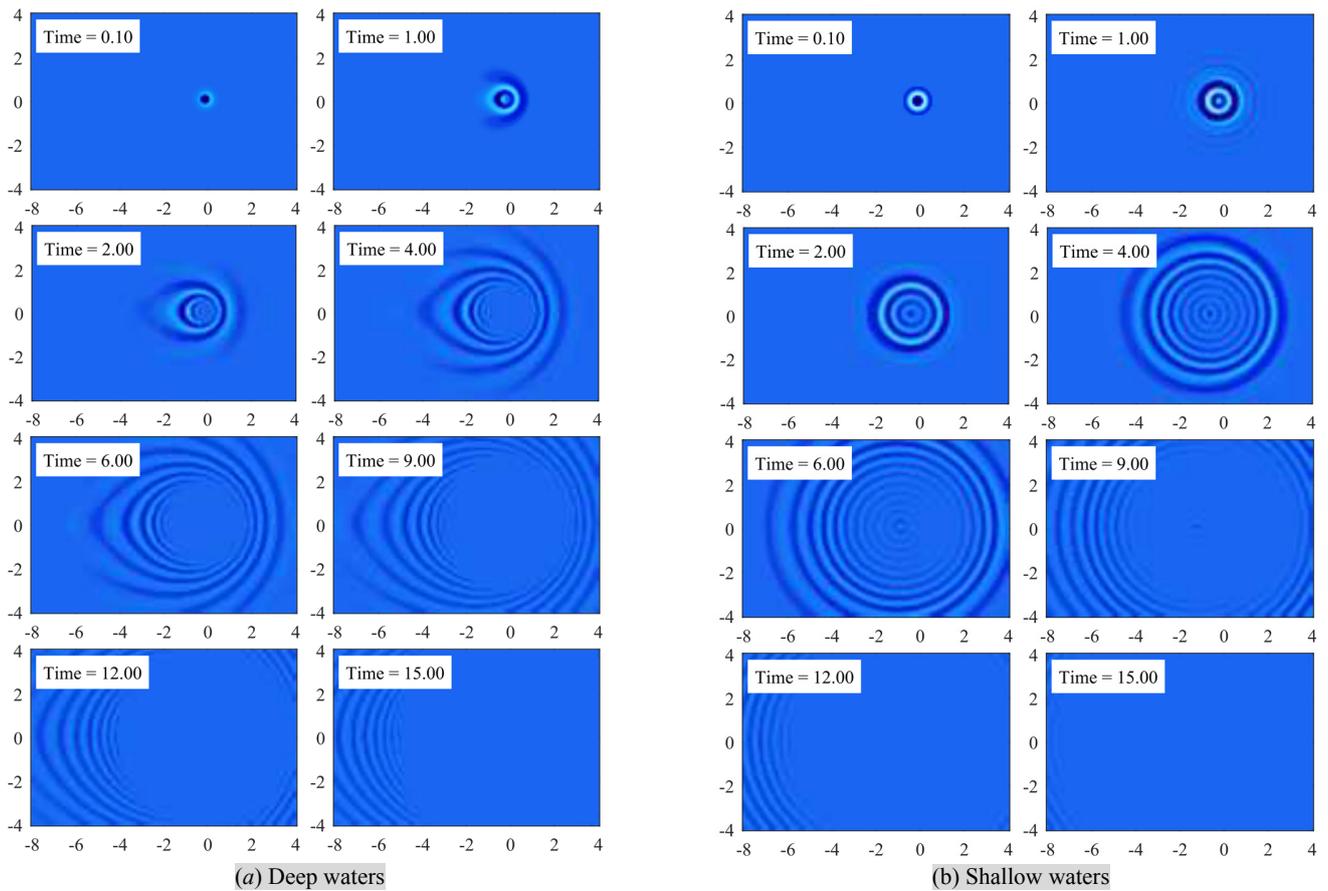

Figure 6. Waves of deep and shallow waters at different times $T$. In all panels $Fr_s$=1, $Fr$=0, and $PI$=2e-4. Left 8 panels : $H$=20, right 8 panels: H=0.05. Areas in this and all below figures where $\zeta > 0.2\zeta_{max}(X,Y,T=0.1)$ are white, $\zeta < -0.2\zeta_{max}(X,Y,T=0.1)$ are black, with linear color gradient for amplitudes in between.

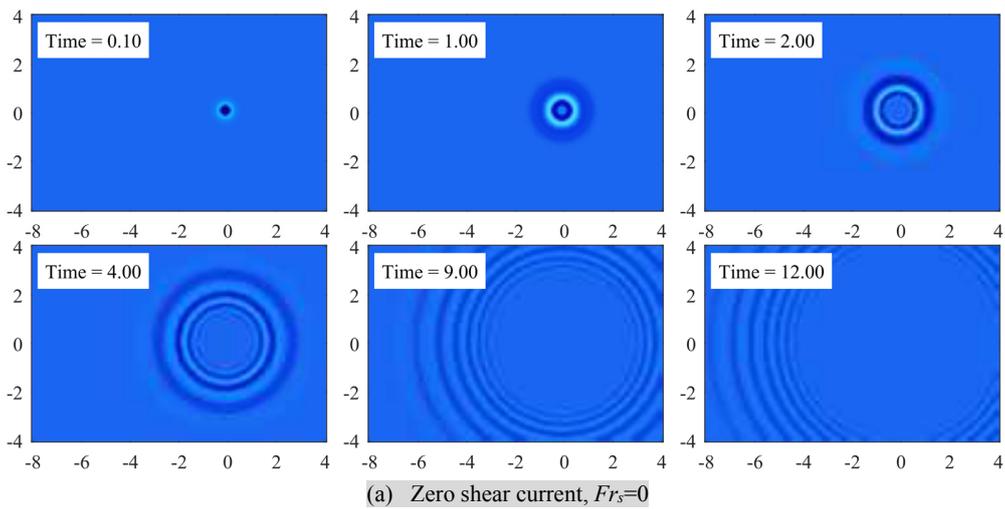

(a) Zero shear current, $Fr_s$=0

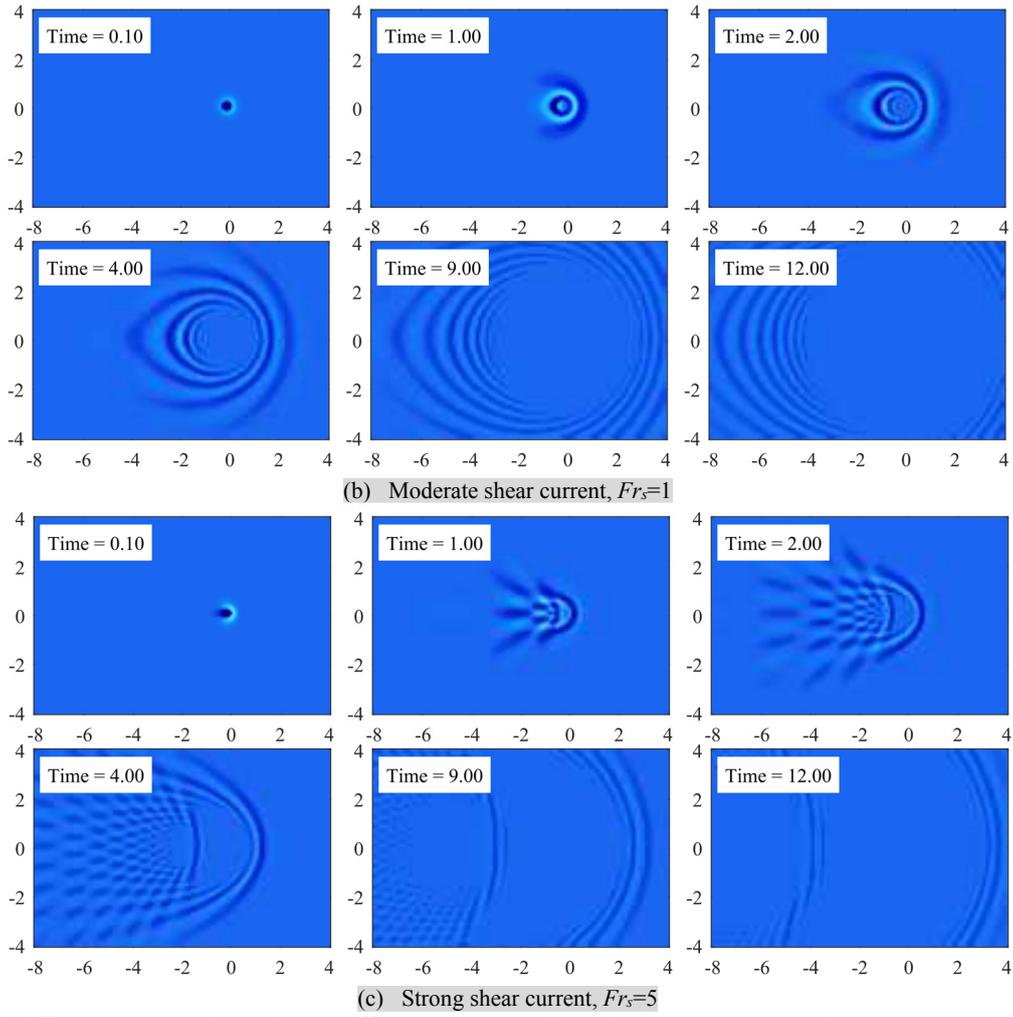

(b) Moderate shear current, $Fr_s$=1

(c) Strong shear current, $Fr_s$=5

Figure 7. Waves with different shear Froude numbers at different times $T$. In all panels: $Fr_{U_0}$=0, PI=2e-4 and $H$=20. Color scaling is as in figure 6.

(a) $β=180°$, $Fr$=0.5, $Fr s$=0.8, $H$=10

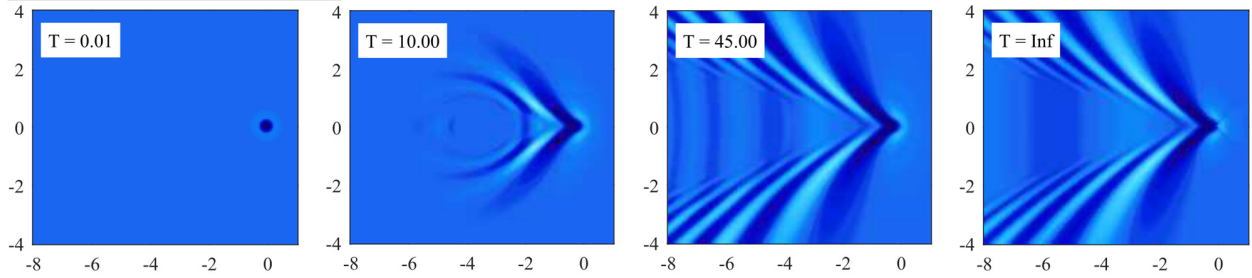

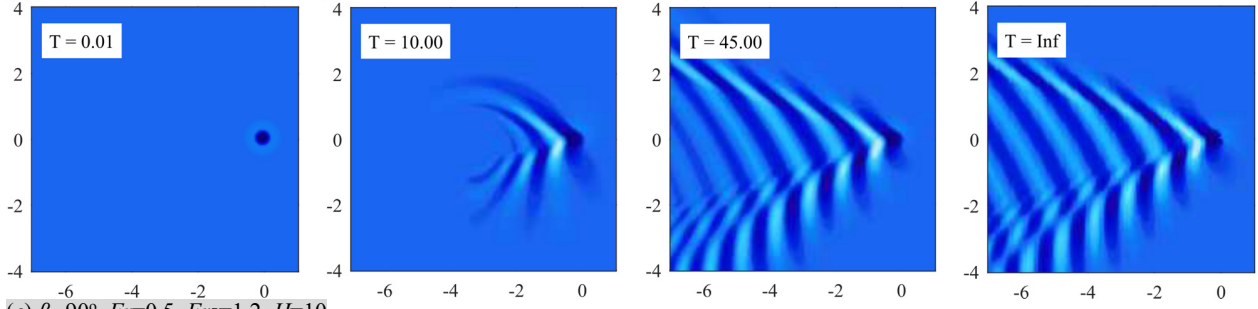

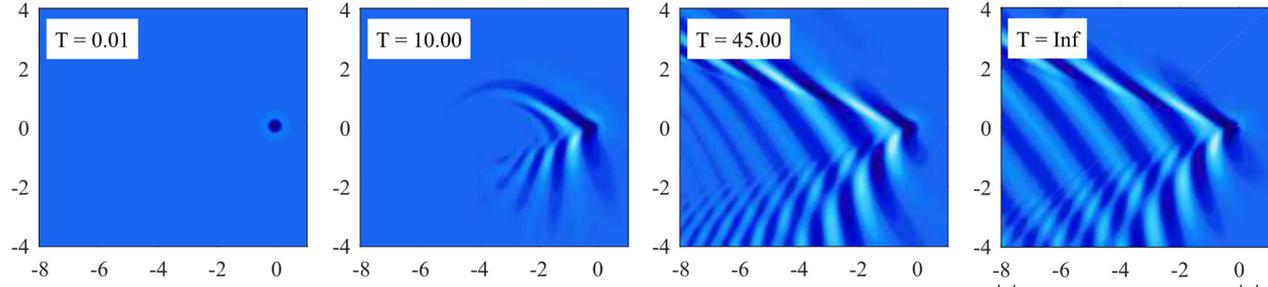

Figure 8. Wave patterns at different times under various circumstances. In all panels, areas where $\zeta > 0.6|\zeta|_{max}$ are white, $\zeta < -0.6|\zeta|_{max}$ are black, with linear color gradient for amplitudes in between; and the axis is scaled by $\Lambda = 2\pi Fr^2$.

*Waves from a steady pressure distribution at different times*

Fig.8 presents the wave patterns resulted from the steady pressure distribution from $T = 0.01$ to $T = \infty$. Calculations were performed using Eq. 25, and no radiation was necessary for the numerical results of Fig.8. The transition from an initial ring wave into a ship wave is obvious, starting from the initial input, and gradually proceeding to the steady waves which is shown in the figures at '$T=Inf$'. Moreover, asymmetry due to the presence of side-on shear current is conspicuous in Fig. 8-b and Fig. 8-c, as reported and discussed previously by Ellingsen (2015b) and Li & Ellingsen (2015a,b). It can be seen from Fig. 8-b and Fig. 8-c that larger vorticity results in more obvious asymmetry of wave patterns. This accords well to the previous results. In comparing the steady wave patterns of Fig. 8 at $T=Inf$ with those of Ellingsen (2015b) and Li & Ellingsen (2015a,b), note that the defintion of $\beta$ differs from that in those papers by 180º due to the difference in coordinate system (those references have the water surface with the pressure source moving, here it is *vice versa*).

Fig.9 presents wave patterns of different times with $Fr=0$. Apparently, according to Fig.9, the transient waves vanish at large times. Moreover, it further indicates that, when there is no current velocity rising at free surface, there is no steady wave that exists. Under real situation, it further implicates that a still 'ship' depressing the free surface will not result in steady waves even there is linearly shear current beneath the free surface with zero surface velocity.

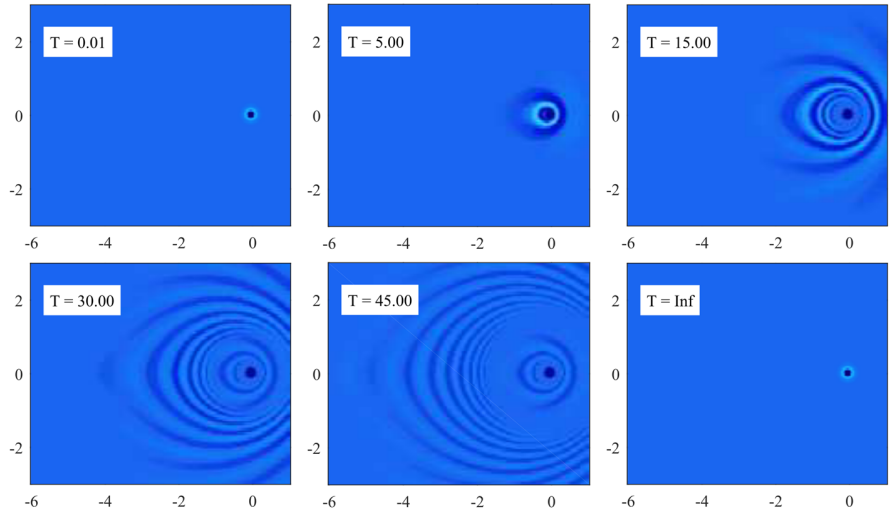

Figure 9. Wave patterns at different times with $Fr=0$, $Frs=0.8$, $H=10$. In all panels, areas where $\zeta > 0.2|\zeta|_{max}$ are white, $\zeta < -0.2|\zeta|_{max}$ are black, with linear color gradient for amplitudes in between; and the axis is scaled by $\Lambda = 2\pi Frs^2$.

## WAVE RESISTANCE & ROLE OF THE TRANSIENT WAVES

We study the effects of the transient waves further by deriving the expressions for wave resistance. Generally, the wave resistance is expressed as

$$R = \int d^2\mathbf{r}\, p_{ext}(\mathbf{r})(V^{-1}\mathbf{V}\cdot\nabla)\zeta(\mathbf{r}) \qquad (48)$$

in which $\mathbf{V}$ is the velocity vector of the moving source, here $-U_0$ is $V$ as it is the relative moving speed of the steady pressure, $\nabla = (\partial/\partial x, \partial/\partial y)$.

Substituting Eq. 33 to Eq. 48 yields

$$\begin{cases} R^t(\mathbf{r},t) = -i\iint \dfrac{d^2 k}{(2\pi)^2} \dfrac{\widetilde{p_{ext}}^2 k^2 \cos\gamma \tanh kh}{2\rho\sqrt{\omega_1^2+\omega_2^2}} \left(\dfrac{e^{-i\omega_+ t}}{\omega_+ - i\varepsilon} - \dfrac{e^{-i\omega_- t}}{\omega_- - i\varepsilon}\right) \\ R^s(\mathbf{r},t) = i\iint \dfrac{d^2 k}{(2\pi)^2} \dfrac{\widetilde{p_{ext}}^2 k^2 \cos\gamma \tanh kh}{2\rho\sqrt{\omega_1^2+\omega_2^2}} \left(\dfrac{1}{\omega_+ - i\varepsilon} - \dfrac{1}{\omega_- - i\varepsilon}\right) \end{cases} \qquad (49)$$

From Eq. 49, it can be observed that the major difference between the transient and the steady wave resistance lies in the time dependent trigonometrical function $\exp(i\omega_+ t)$ and $\exp(i\omega_- t)$ in the integrand of the transient wave resistance.

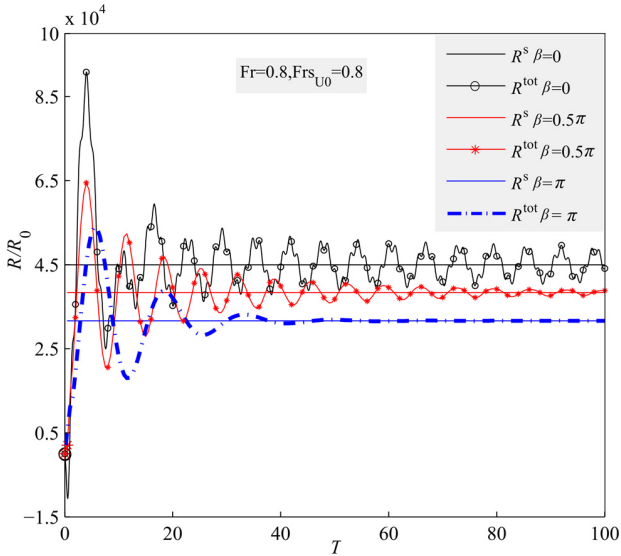

Figure 10. Wave resistances at different times. In the figure, $R_0 = p_0^2/(4\pi^3\rho g)$.

A similar analysis can be made as we have made in previous subsection regarding how transient waves vanish at large times. The transient wave resistance behaves in a similar manner at large times as the transient wave patterns. It ultimately vanishes when time is large enough, which is illustrated in Fig.10. Fig.10 presents total wave resistance (including transient and steady part) and the steady wave resistance alone, at different times for different $\beta$. Striking differences between different directions of ship motion, $\beta$, can be clearly observed. The existence of the vorticity has significant effects on the decay of the transient wave pattern – transient waves decay much faster for $\beta=\pi$ (ship motion along the current) than for $\beta=0$ (ship motion against the current). Physically, gravity waves running against the current can have group velocity only slightly lower than phase velocity (see Ellingsen 2014a). Hence the transverse waves behind a ship going against the current are aided along by the sub-surface vorticity, and transients remain longer in the ship's near-field where they can contribute to resistance, before being left behind. Conversely, waves running along the current can have phase velocity far greater than group velocity, and transients are left behind very quickly.

The fact that transient waves can take very long to leave a ship's near field for motion against the current means an increased chance of significant effects from accumulation of transient effects, e.g., due to maneuvering. This could present increased seakeeping challenges when running against a shear flow, compared to a uniform flow. We intend to investigate this phenomenon further in future work.

## CONCLUSIONS

In this paper we obtained a solution to the problem of linear surface waves due to a general, time-dependent applied pressure distribution, in the presence of a sub-surface shear current with uniform vorticity. Solutions are obtained in both finite and infinite water depth. As examples, corresponding analyses are carried out for two special cases: a short pressure impulse at $t=0$, and a suddenly appearing, constant pressure distribution.

By numerical simulations we showed how the impulsive pressure, which gives rise to an initial velocity distribution in the fluid, creates a ring wave surface elevation pattern at subsequent times, and that our solution conforms with previously known results for the Cauchy-Poisson problem. An effect of the sub-surface shear current of uniform vorticity is to create asymmetry of the wave motion between upstream and downstream directions, a direct result of the non-isotropic dispersion relation. Moreover, we confirmed a previous observation that vorticity of the current has more prominent influence in deep waters.

We next studied a steady pressure distribution which appears suddenly at $t=0$, a transient situation which begins as a ring wave and develops, if the water surface is moving relative to the pressure distribution, into the ship wave pattern studied previously by Ellingsen (2015) and Li & Ellingsen (2015a). We showed that the wave solution splits naturally into a stationary and a transient term where the stationary contribution can be identified as the ship wave solution, and the transient term can be shown to eventually vanish at large times. We specifically calculated the time-dependent wave resistance on such a suddenly appearing steady wave source. In the presence of a shear current, the time that takes for the transient contribution to vanish was found to depend greatly on the ship's direction of motion. For a ship moving against the direction of the shear current, transient effects can take a long time to vanish because phase velocity only slightly exceeds group velocity, and effects of transient waves, e.g., from maneuvering, can easily accumulate, potentially posing additional seakeeping challenges compared to a uniform current.